\documentclass[%
 reprint,
superscriptaddress,
 amsmath,amssymb,
 aps,
prl,
]{revtex4-2}

\usepackage{graphicx}
\usepackage{dcolumn}
\usepackage{bm}

\renewcommand{\selectlanguage}[1]{}
\begin{document}

\preprint{APS/123-QED}

\title{Uncovering the origin of interface stress enhancement and compressive-to-tensile stress transition in immiscible nanomultilayers}

\author{Yang Hu}
\affiliation{Laboratory for Advanced Materials Processing, Empa - Swiss Federal Laboratories for Materials Science and Technology, Thun, Switzerland\\}
\affiliation{National Centre for Computational Design and Discovery of Novel Materials MARVEL, Empa, Thun, Switzerland\\}%

\author{Giacomo Lorenzin}%
\affiliation{Laboratory for Joining Technologies and Corrosion, Empa - Swiss Federal Laboratories for Materials Science and Technology, Duebendorf, Switzerland\\}%

\author{Jeyun Yeom}%
\affiliation{Laboratory for Joining Technologies and Corrosion, Empa - Swiss Federal Laboratories for Materials Science and Technology, Duebendorf, Switzerland\\}%

\author{Manura Liyanage}%
\affiliation{Laboratory for Advanced Materials Processing, Empa - Swiss Federal Laboratories for Materials Science and Technology, Thun, Switzerland\\}%
\affiliation{National Centre for Computational Design and Discovery of Novel Materials MARVEL, Empa, Thun, Switzerland\\}%

\author{William A. Curtin}%
\affiliation{School of Engineering, Brown University, Providence, RI 02906 USA\\}%
\affiliation{National Centre for Computational Design and Discovery of Novel Materials MARVEL, EPFL, Lausanne, Switzerland\\}%

\author{Lars P.H. Jeurgens}
\affiliation{Laboratory for Joining Technologies and Corrosion, Empa - Swiss Federal Laboratories for Materials Science and Technology, Duebendorf, Switzerland\\}%

\author{Jolanta Janczak-Rusch}
\affiliation{Laboratory for Joining Technologies and Corrosion, Empa - Swiss Federal Laboratories for Materials Science and Technology, Duebendorf, Switzerland\\}%

\author{Claudia Cancellieri*}
\affiliation{Laboratory for Joining Technologies and Corrosion, Empa - Swiss Federal Laboratories for Materials Science and Technology, Duebendorf, Switzerland\\}%
\email{claudia.cancellieri@empa.ch}

\author{Vladyslav Turlo*}
\affiliation{Laboratory for Advanced Materials Processing, Empa - Swiss Federal Laboratories for Materials Science and Technology, Thun, Switzerland\\}%
\affiliation{National Centre for Computational Design and Discovery of Novel Materials MARVEL, Empa, Thun, Switzerland\\}%
\email{vladyslav.turlo@empa.ch}

\date{\today}

\begin{abstract}
The intrinsic stress in nanomultilayers (NMLs) is typically dominated by interface stress, which is particularly high in immiscible Cu/W NMLs. Here, atomistic simulations with a chemically-accurate neural network potential reveal the role of interfacial intermixing and metastable phase formation on the interface stress levels.  These results rationalize an experimentally-reported compressive-to-tensile transition as a function of NML deposition conditions and the extremely high interface stresses under some conditions.
\end{abstract}

\keywords{interface stress, Cu/W nanomultilayers, molecular statics, intermixing, metastable phases}

\maketitle

Nanomultilayers (NMLs) are composed of stacks of alternating nanolayers of two or more dissimilar materials or phases. The NML microstructure gives rise to unique properties that are dominated by the high volumetric density of the hetero-interfaces \cite{Trevizo2020,clemens1999structure}. Literature studies report NMLs with outstanding magnetic, optical, mechanical, and radiation tolerance properties, leading to numerous applications in micro-/nanojoining, microelectronics, micro-/nanoelectromechanical systems, optical coatings, plasma research, etc. \cite{Trevizo2020}. 

NMLs are commonly fabricated by physical vapor deposition techniques, in particular magnetron sputtering that offers great precision at relatively fast deposition speeds \cite{Trevizo2020}. However, the deposition of an NML on a parent substrate leads to the generation of intrinsic growth stresses that can affect mechanical \cite{Abadias2007}, thermal \cite{LorenzinShafkat2022}, and atomic transport \cite{Troncoso2024} properties.  While the underlying mechanisms of intrinsic stress generation in thin-film systems are well understood \cite{Abadias2018}, the intrinsic stress generation mechanisms in NML systems with their intrinsically high density of buried interfaces are less well studied and documented. In particular, there is a lack of fundamental knowledge and engineering know-how for tailoring the interface stress contributions during deposition/fabrication of NMLs. 

The evolution of the average growth stress during thin-film and NML deposition is commonly measured in real-time using the wafer curvature technique \cite{Ruud1993}. The final depth-dependent residual stress gradient in the as-deposited NML/substrate assembly \cite{cancellieri2021strain} arises from the superposition of bulk residual stresses (due to e.g. ion implantation), interface stresses at successive phase boundaries (hetero-interfaces), and residual stresses at high-angle grain boundaries. For most NML systems, the stress contribution from grain boundaries can generally be neglected relative to that from the hetero-interfaces \cite{Birringer2009}. 
After NML deposition, the average residual stresses in the individual phases of the as-deposited NML can be assessed \textit{ex situ} by X-ray diffraction (XRD) \cite{LorenzinShafkat2022}.  The interface stress per layer thickness can then be estimated from the measured difference between the average total intrinsic stress, as derived from the substrate curvature, and the average residual stresses in each phase in the NML \cite{Lorenzin2024, Ruud1993}.  

\begin{figure}
\includegraphics[width=0.45\textwidth,clip,trim=0cm 0.3cm 0cm 1cm]{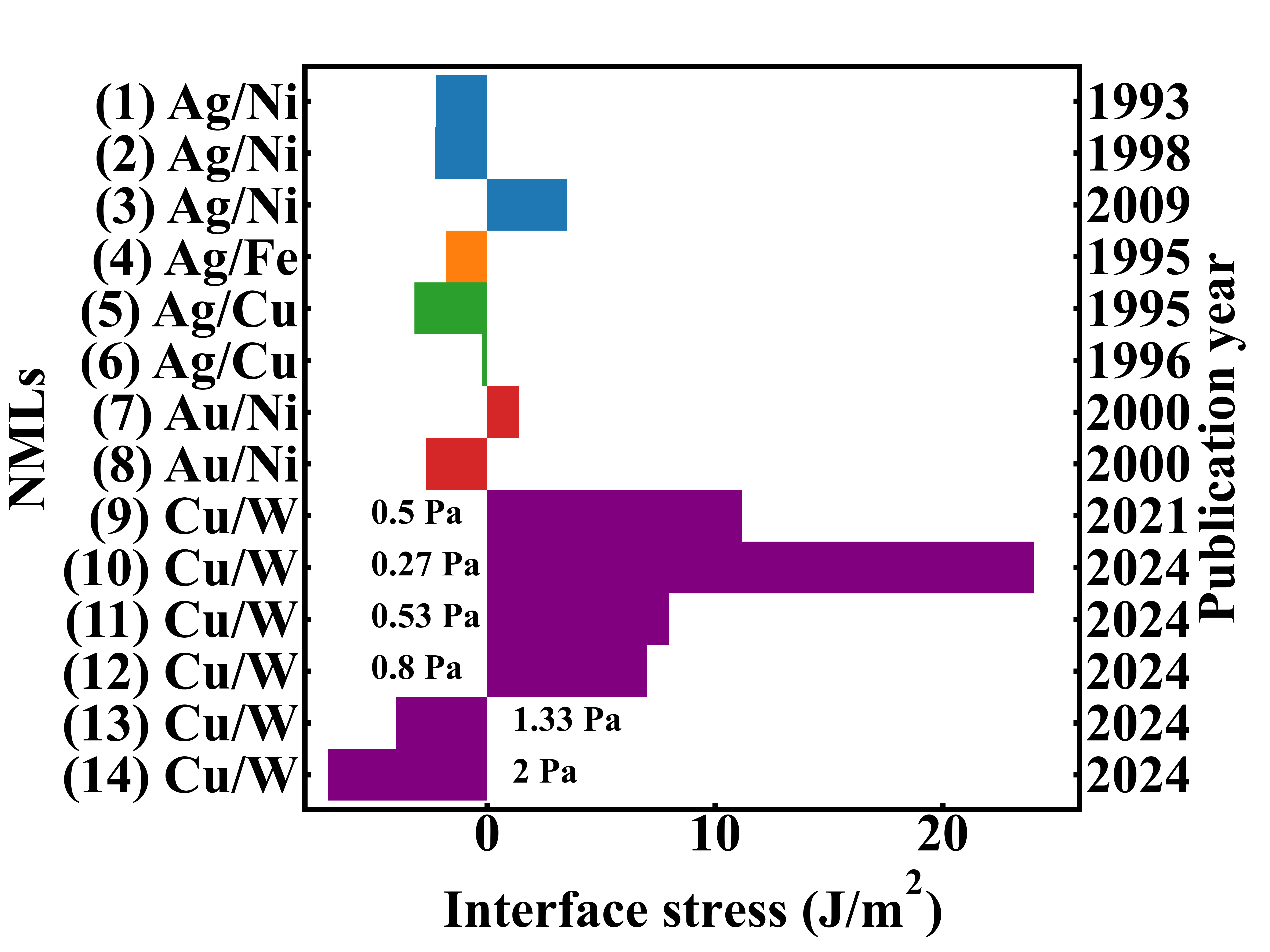}
\caption{\label{fig1} Reported interface stresses for different as-deposited NMLs: (1) (111) Ag/Ni \cite{Ruud1993}; (2) (111) Ag/Ni \cite{Schweitz1998}; (3) untextured Ag/Ni \cite{Birringer2009}; (4) (100) Ag/Fe \cite{Scanlon1995}; (5) (111) Ag/Cu \cite{Berger1995}; (6) (111) Ag/Cu \cite{Shull1996}; (7) (111) Au/Ni \cite{Schweitz2000}; (8) (111) Au/Ni \cite{Labat2000}; (111) Cu/(110) W at different Ar pressures of (9) 0.5 Pa \cite{Druzhinin2021}, (10) 0.27 Pa, (11) 0.53 Pa, (12) 0.8 Pa, (13) 1.33 Pa, and (14) 2 Pa \cite{Lorenzin2024}.}
\end{figure}

Interface stresses reported for many NMLs are summarized in Fig. \ref{fig1}, and can be positive or negative, even for the same NML system (see e.g. Ag/Ni, Au/Ni, and Cu/W in Fig. \ref{fig1}). Notably, the interface stresses for face-centered-cubic (\textit{fcc}) Cu/body-centered-cubic (\textit{bcc}) W NMLs are substantially higher than in other NML systems, also exhibiting both positive and negative values. Since Cu and W are chemically immiscible and the interface stress levels in Cu/W NMLs are relatively high, the Cu/W system is considered as an ideal model system for studying interface stress-related phenomena in NMLs.  
Here, we show that the magnitude and compressive-to-tensile transition of intrinsic stress in Cu/W NMLs can be attributed to variations in Cu-W intermixing at the interface as a function of the Ar pressure during magnetron sputtering (Fig. \ref{fig1}).  Intermixing also rationalizes the reduction of interface stress with increase in annealing temperature due to interfacial demixing.  Analysis leads to the proposal that the highest positive values of interface stress are due to the formation of a Cu-rich \textit{bcc} phase at the interface, a result confirmed by here experiments.

The interface stress $f(\varepsilon)$ under a biaxial in-plane strain, $\varepsilon$, is related to the interface energy, $\gamma(\varepsilon)$, by the Shuttleworth equation
\begin{equation} 
f(\varepsilon)=\gamma(\varepsilon)+\frac{\partial{\gamma(\varepsilon)}}{\partial{\varepsilon}}
\label{eq:interfstre}
\end{equation} 
The value of $f(\varepsilon)$ can thus be determined by computing interface energy versus biaxial strain, as recently performed by density-functional theory (DFT) calculations for idealized Cu/W interface structures \cite{Lorenzin2024}. The DFT calculations predict an evolution of the interface stress from positive to negative in Cu/W NMLs, in qualitative agreement with experiments. However, the calculated interface stress values are much smaller than experiments \cite{Lorenzin2024}. This quantitative discrepancy was attributed to the limited DFT system size of a few hundred atoms, which precludes exploration at experimental nanolayer thicknesses and does not account for any distribution of crystallographic orientation relationships (CORs) along the Cu/W interface (since nanocrystalline Cu and W phase typically exhibit a random in-plane texture).  The limitations of DFT can be partially mitigated by the development and application of a machine learning interatomic potential \cite{Lorenzin2024, mishin_machine-learning_2021,seko_tutorial_2023}. Thus, here, a new machine-learned, DFT-trained, neural network (NN) interatomic potential for the Cu-W system is used \cite{Liyanage2024} to execute molecular statics simulations at realistic size scales and for a variety of CORs and local structures.  The Cu-W NN potential reproduces the Cu and W elastic moduli, generalized stacking fault energies for common slip systems, Cu/W interface energies for common CORs, substitutional energies for Cu in W and W in Cu, as well as the formation energies of metastable solid solutions and intermetallic phases. The potential has already been successfully applied to explain the drop in experimental Young's modulus of Cu/W NMLs due to the excess free volume introduced during the fabrication process \cite{Lorenzin2024preprint}.   

\begin{figure}
\includegraphics[width=0.47\textwidth,clip,trim=2.8cm 2.5cm 2.5cm 2cm]{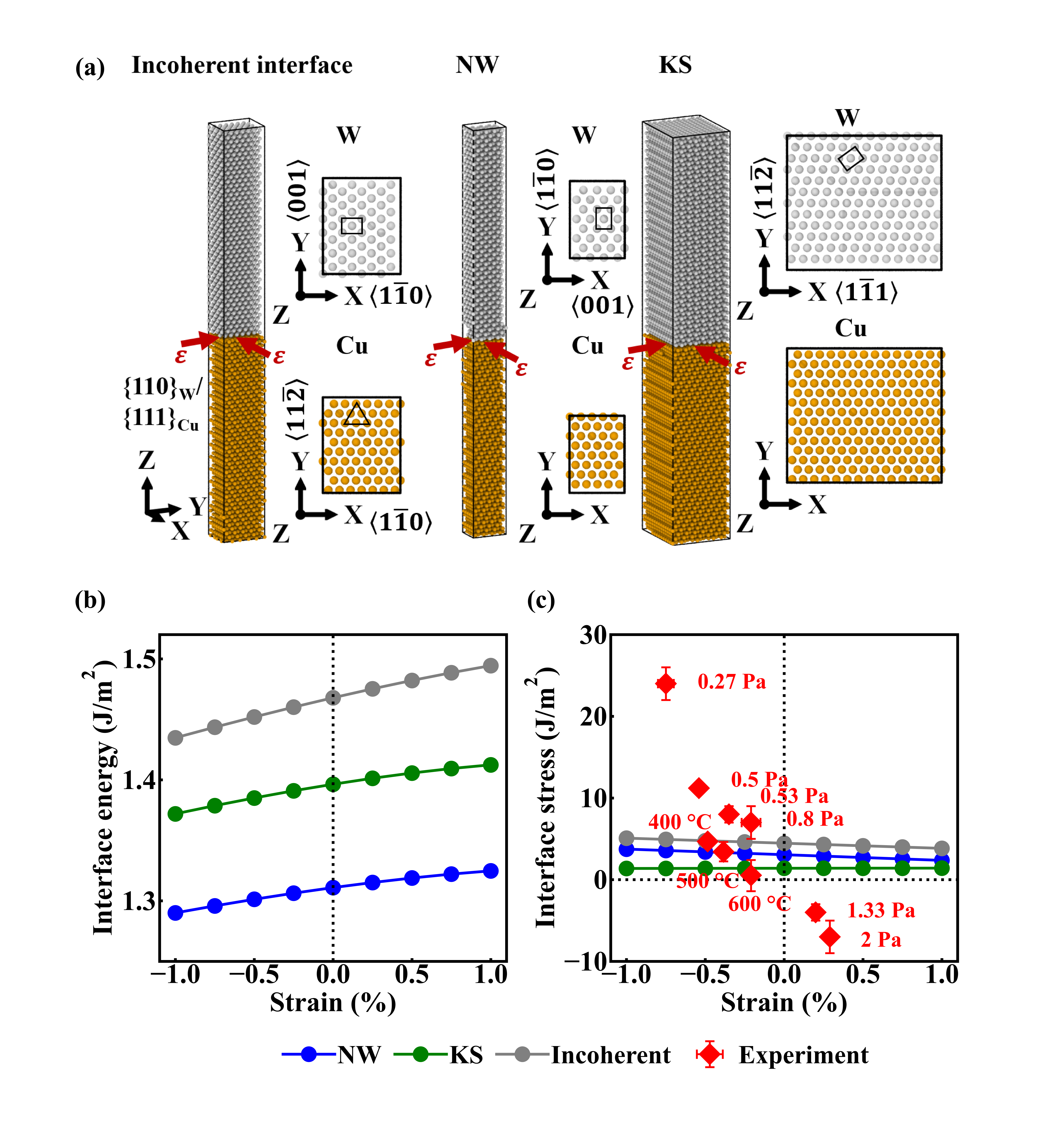}
\caption{\label{fig2} (a) MD simulation cells for three $\{110\}_{W}/\{111\}_{Cu}$ interfaces with different in-plane orientations. Cu atoms are in gold color and W atoms are grey. (b) Interface energy and (c) interface stress versus in-plane strain.}
\end{figure}

Molecular statics simulations were performed in periodic Cu/W NMLs while adopting the experimentally-observed and theoretically-predicted out-of-plane $\{110\}_{\textit{bcc}}/\{111\}_{\textit{fcc}}$ COR \cite{Cancellieri2016, HALL1972257, Girault2006, Druzhinin2019,moszner2016nano}.  Three different in-plane CORs with atomically sharp Cu/W interfaces were studied:(\textit{i}) an incoherent interface (Figure \ref{fig2}(a)), constructed using CellMatch \cite{Lazic2015} to minimize the residual lattice mismatch to 0.3\% relative to the T=0K lattice constants of Cu and W, and the common (\textit{ii}) Kurdjumov-Sachs (KS) and (\textit{iii}) Nishiyama-Wasserman (NW) CORs, constructed as described in Ref. \cite{Lorenzin2024} with misfit strains of 0.245\% and 1.022\%, respectively.
Nanolayer thicknesses of 10 nm (i.e. comparable to experimental layer thickness) were studied. Displacements corresponding to biaxial in-plane strains ($\varepsilon=\varepsilon_{xx}=\varepsilon_{yy}$) were applied. The simulation cell was then relaxed along the normal (Z) axis to zero pressure. Calculations were performed using the NN as implemented in the n2p2 code \cite{Singraber2019} within the Large-scale Atomic/Molecular Massively Parallel Simulator (LAMMPS) \cite{Plimpton1995}. The resulting interface energy was computed from the total system energy $E_{tot}^{bl}(\varepsilon)$ as
\begin{equation}    
\gamma(\varepsilon)=\frac{E_{tot}^{bl}(\varepsilon)-N_{\text{Cu}}^{bl}\cdot e_{\text{Cu}}^{b}(\varepsilon)-N_{\text{W}}^{bl}\cdot e_{\text{W}}^{b}(\varepsilon)}{2A(\varepsilon)},
\label{eq:interfen}
\end{equation}
where $e_{\text{Cu}}^{b}(\varepsilon)$ and $e_{\text{W}}^{b}(\varepsilon)$ are the energies per atom of bulk Cu and W atoms under the applied in-plane strain, $N_{\text{Cu}}^{bl}$ and $N_{\text{W}}^{bl}$ are the total numbers of Cu and W atoms in the bilayer, and $2A$ is the total interface area in the periodic simulation cell.

Figure \ref{fig2}(b) shows the interface energies versus in-plane strain for the three different in-plane orientations.  As expected, the NW orientation has the lowest interface energy and the incoherent interface has the highest interface energy, although the differences are less than 200 mJ/m$^2$.  The interface energies for NW and KS at zero strain are comparable to the corresponding DFT values \cite{Bodlos2022,Liyanage2024,Lorenzin2024}.  The interface energy versus in-plane strain was fitted to a parabola, $\gamma=a\varepsilon^2+b\varepsilon+c$ and the corresponding interface stresses derived using Eq. \ref{eq:interfstre}, as shown in Figure \ref{fig2}(c). For all three in-plane CORs, the interface stresses show only a weak dependence on in-plane strain, are always positive, and have magnitudes much lower than those experimentally determined for as-deposited Cu/W NMLs; these results are in line with our previous DFT and NN potential calculations \cite{Lorenzin2024}. Strikingly, the predicted interface stresses do agree very well with the interface stress values measured after post-annealing at different temperatures \cite{Druzhinin2021}.  This points to intermixing at the interface as an explanation for the interface stresses in as-deposited Cu/W NMLs because Clemens and Ramaswamy \cite{Clemens2000} highlighted that the interface stress must be extremely sensitive to small amounts of intermixing, and intermixing should be reduced by low-temperature annealing.  Specifically, post-annealing after sputter deposition will reduce the non-equilibrium defect density from the sputtering process and also induce grain rotation and grain growth in the confined Cu nanolayers, thereby allowing the development of more coherent interfaces with lower energy and stress.  Since the predicted interface stresses in Cu/W NMLs with atomically sharp (i.e. unmixed) interfaces show a relatively weak dependence on the in-plane COR and much smaller magnitudes than experiments on as-deposited Cu/W NMLs, we study the consequences of mechanical and/or chemical intermixing at the Cu/W interfaces as a likely explanation for the discrepancy between theory and experiment. 

\begin{figure}
\includegraphics[width=0.47\textwidth,clip,trim=4.3cm 3.9cm 3cm 1.8cm]{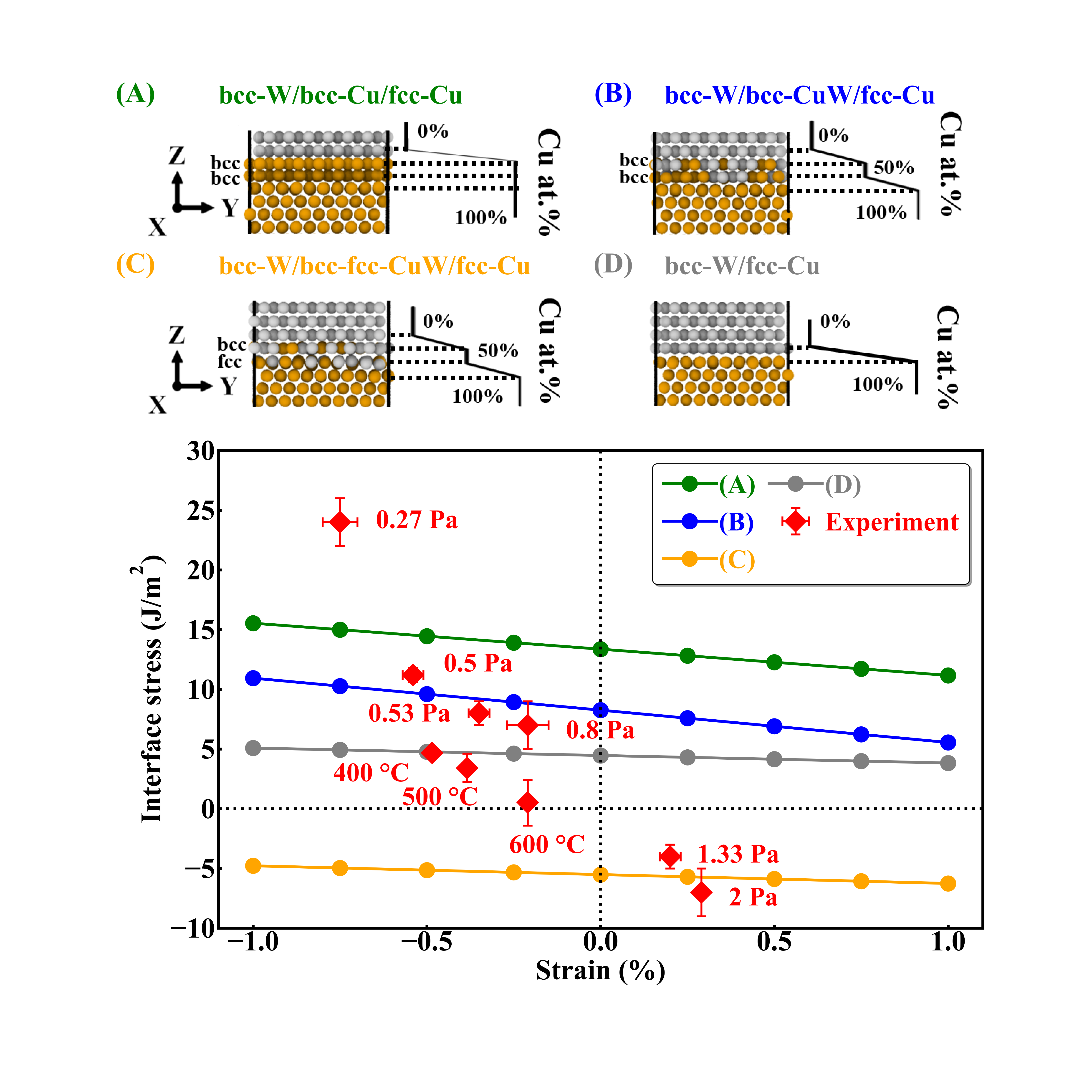}
\caption{\label{fig3} Atomic snapshots of incoherent interfaces with varying structure and chemistry of interfacial layers (top) and their corresponding interface stresses with respect to in-plane strains (bottom). 
}
\end{figure}

Although Cu and W are immiscible, magnetron sputtering is a non-equilibrium deposition process and will thus induce some degree of mechanical intermixing at the Cu/W interfaces during the alternating W and Cu deposition steps. 
Mechanical intermixing should only occur within the first few atomic planes at the interface, being more pronounced towards higher argon pressures and at ambient temperatures, which reduce the thermal energy of the deposited adatoms and thus limit atomic surface and interface mobility and thereby chemical demixing.  To test this hypothesis quantitatively, we use the incoherent interface and randomly exchange Cu and W atom types within several layers around the interface. 
Although the composition of such intermixed layers has never been reported for Cu/W NMLs, Guo and Thompson \cite{Guo2018} quantified the intermixing in immiscible 5nm/5nm Cu/V NMLs using atom probe tomography and reported a 2-nm-thick intermixed interfacial region with a 50-50 at.\% composition. Accordingly, we investigate mechanical intermixing over two atomic layers (one fcc and one bcc) with a 50-50 at.\% Cu-W composition. The computed interface stress is shown in Fig. \ref{fig3}(orange line) and indeed intermixing of Cu and W in just two interfacial planes not only reverses the sign of interface stress but also matches the experimentally-measured magnitudes for high argon pressures. 

In contrast, for low argon pressures, the thermal energy of the deposited heavy W adatoms can enhance the surface diffusion of Cu and lead to so-called surface-energy-driven intermixing \cite{Clemens2000,Li2023,reichel2006modeling}.  Surface energies of W are typically twice those of Cu surfaces \cite{Liyanage2024}, suggesting that it is energetically favorable for Cu atoms to diffuse over the W adatoms, effectively burying W atoms during the W nanolayer deposition steps \cite{Villain2007}. The resulting intermixed Cu-W planes most likely adopt the \textit{bcc} structure, which has lower energy than random \textit{fcc} Cu-W up to ~70 at.\% Cu \cite{Liang2017,Liyanage2024}. 
Accordingly, the interface stress in an NML with two \textit{bcc} intermixed layers with 50 at.\% Cu has been simulated, as shown in Fig. \ref{fig3}(blue line). 
Indeed, the interface stress increases substantially relative to all interfaces without atomic intermixing. This supports our hypothesis of surface-energy-driven chemical intermixing at low argon pressure.  

Our results show that slight intermixing within just two atomic planes explains the majority of experimental data except for the extremely high positive interface stress in Cu/W NMLs deposited under the lowest Ar pressure. This suggests a careful analysis of Cu adatom deposition on W surfaces without thermodynamically favored intermixing.  There is evidence that highly mobile Cu adatoms may diffuse into W grain boundaries (in accordance with the well-accepted theory of compressive stress generation during the post-coalescence film growth \cite{chason2002origin,chason2016tutorial}) and form two pseudomorphic \textit{bcc} Cu planes that are stabilized by W surfaces at the opposite side of a W grain boundary groove \cite{BAUER1984}.  Accordingly, unmixed \textit{fcc/bcc} interfaces with two atomic planes of pure \textit{bcc} Cu at the interface were simulated.  As expected, two layers of pure \textit{bcc} Cu at the interface are indeed stable (at least at 0 K) and even result in a lower interface energy as compared to other intermixed interface structures, see Supplementary Table II.  This is possibly due to the high-energy penalty associated with mutual intermixing as compared to the \textit{fcc-bcc} energy difference for Cu \cite{Liyanage2024}).
The corresponding interface stresses for two planes of \textit{bcc} Cu are shown in Fig. \ref{fig3}(green line) with magnitudes reaching ~15 J/$\mathrm{m}^{2}$, approaching the highest experimentally measured interface stress values. 

Our results suggest that the highest interface stresses are achieved by the formation of a metastable \textit{bcc}-structured Cu-rich phase at the Cu/W interface. Experimental validation of such a phenomenon by e.g. cross-sectional High-Resolution Transmission Electron Microscopy (HR-TEM) and/or XRD is quite challenging for a 10 nm-nanolayer thickness and with the intrinsic microscopic waviness of buried Cu/W interfaces. That is, any TEM micrograph represents an edge-on projection along the TEM lamellae thickness. XRD investigations would mainly probe the bulk equilibrium phases and not resolve the tiny contribution from the interface structures. A substantial reduction of the nanolayer thickness amplifies the volumetric density of the interface material, which may help to experimentally examine the interface structure.
Therefore, a 2 nm/2 nm Cu/W NMLs was fabricated at the same Ar pressure of 0.27 Pa at which 10 nm NMLs show a very high interface stress; details of the fabrication and characterization are provided in the Supplementary Materials. 

The scanning transmission electron microscopy (STEM) image in Fig. 4(a) shows a cross-sectional view with a clear layered structure of the Cu/W NML with diffuse interfaces between the alternating Cu and W nanolayers. Upon closer inspection of the atomic structure of individual layers with HR-TEM, \textit{bcc}-structured columnar grains spanning across the layers can be observed (see Fig. 4(b)). The inset of Fig. 4(b) shows the fast Fourier transformation of the area marked by the red rectangle, with a Fourier filter applied to Fig. 4(b) to reduce the noise level, as well as to identify the atomic arrangement clearly. The thus-obtained pattern is a typical \textit{bcc} lattice and is used to produce Fig. 4(c) through inverse fast Fourier transformation. Figs. 4(d,e) show X-ray diffraction pole figures for the \{110\} W planes \cite{Parrish:a02923} and the \{111\} Cu planes \cite{GUPTA2018239}. The pole figures reveal that both W and Cu layers display a fiber texture, i.e., they have a preferential out-of-plane direction, but the grains are randomly oriented in-plane. The ring in the pole figure of W (Fig. 4(d)) has its maximum at $\psi = 60^{\circ}$, as expected for a \textit{bcc} polycrystalline sample with a preferential [110] out-of-plane orientation.  For an \textit{fcc} crystal with a preferential [111] out-of-plane orientation, the expected maximum of the ring intensity is $\psi = 70.53^{\circ}$ \cite{Druzhinin2019,Cancellieri2016}, but here it appears at $\psi \approx 67.5^{\circ}$ (Fig. 4(e)). This highlights significant distortions of \textit{fcc} Cu grains due to the formation of \textit{bcc} crystallites at the Cu/W interfaces, as evidenced by the corresponding TEM images in Fig. 4(a-c). Although it is not sufficient enough to claim a complete transformation of the Cu nanolayers into a \textit{bcc}-like structure, the existence of Cu-rich \textit{bcc}-structured nanograins at the Cu/W interfaces is experimentally confirmed.

\begin{figure}
\includegraphics[width=0.45\textwidth,clip,trim=0cm 0.3cm 0.5cm 0cm]{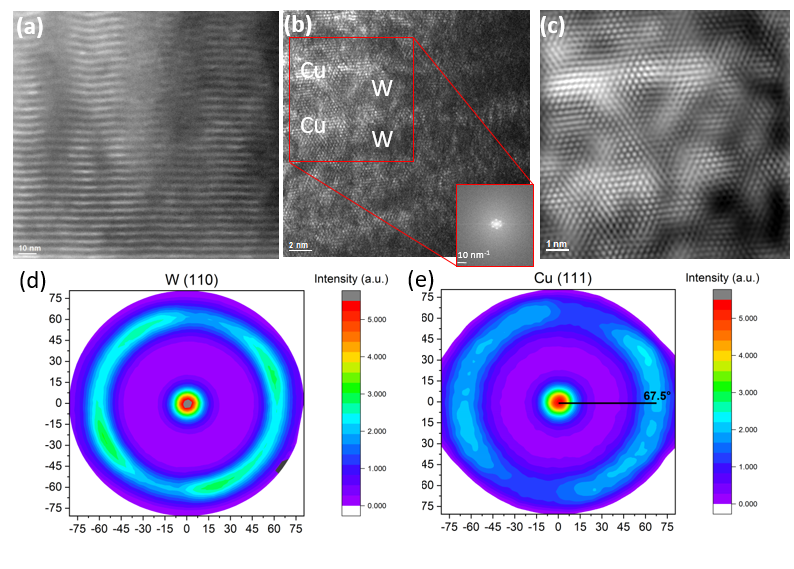}
\caption{\label{fig4} (a) STEM and (b) HR-TEM images of the cross-sectional view of a 2 nm/2 nm Cu/W NML deposited under 0.27 Pa argon pressure. Inset in (b) shows the Fourier Transform of the region inside the red box. (c) HR-TEM image of the inset of (b) improved through noise filtering. Pole figures of (d) the W \{110\}  \cite{Parrish:a02923} and (e) the Cu \{111\} family of planes \cite{GUPTA2018239}. }
\end{figure}

In conclusion, our work demonstrates the critical role of atomic intermixing and metastable phase formation at Cu/W interfaces in enhancing the magnitude of interface stress. Even the smallest amount of intermixing at the interfaces can cause drastic changes in the measured interface stress levels.  Cu incorporation into the \textit{bcc} W nanolayers increases the interface stress, while W incorporation into the \textit{fcc} Cu nanolayers reduces the interface stress and may even invert its sign. This implies that even for highly immiscible systems, like Cu-W, some atomic intermixing at the interfaces may be thermodynamically favorable to reduce the excess Gibbs energy associated with high interface stress levels.  Beyond that, it is demonstrated and confirmed experimentally that the exceptionally high experimental interface stress can be attributed to the formation of Cu-rich metastable phases at the interfaces. We also emphasize the critical, enabling role of our chemically accurate machine-learned potential for studying systems at experimental scales and guiding interpretation of the experimentally measured interface stresses in NMLs.
The fundamental understanding of interface stress phenomena emerging here will help enable the smart design of advanced NML coatings with tunable stress states to control their thermomechanical stability and multifunctional properties.

\begin{acknowledgments}
This research was supported by NCCR MARVEL, a National Centre of Competence in Research, funded by the Swiss National Science Foundation (grant number 205602). G.L. acknowledges the Swiss National Science Foundation (SNSF) under Project No. 200021{\_}192224 for financially supporting this research. 
V.T. thanks the Bern Economic Development Agency for financial support.
\end{acknowledgments}

\bibliographystyle{plain}


\end{document}